\documentclass[10pt]{article}

 \usepackage{amsmath,amsthm,amssymb}
 \usepackage{makeidx,epsfig,lscape}
 \usepackage{color,colortbl}
 \usepackage{fancyhdr}
 \usepackage{xcolor,pict2e}

 \renewcommand{\headrulewidth}{0pt}
 \renewcommand{\footrulewidth}{0.5pt}

 \definecolor{myaqua}{rgb}{0.0,0.5,0.55}
 \definecolor{lightaqua}{rgb}{0.75,0.95,0.95}

 \usepackage[colorlinks = true,
            linkcolor = myaqua,
            urlcolor  = blue,
            citecolor = myaqua]{hyperref}

\usepackage{caption}
\usepackage{floatrow}
\def\lin#1#2{\textcolor[rgb]{0.6,0.6,0.6}{\vspace*{#1mm} \hrule
   height 3 pt \vspace*{#2mm}}}
\def\bt{\begin{tabular}}
\def\et{\end{tabular}}
\def\and{\mbox{ and }}

\def\1{{\bf 1}}

 \def\sectionn#1{\refstepcounter{section}{\color{myaqua}

 \vskip 6mm

 \noindent\Large\bf\thesection. #1}

 \vskip 3mm}

 \def\boxx#1#2#3#4#5{
 {\linethickness{#4pt}\put(#1,#5){\color{myaqua}{\line(1,0){#3}}}}
 \multiput(#1,#2)(0,#4){2}{\line(1,0){#3}}
 \multiput(#1,#2)(#3,0){2}{\line(0,1){#4}}
  }

\begin{document}

 \fancyhead[L]{\hspace*{-13mm}
 \bt{l}{\bf Journal of Applied Mathematics and Physics, 2016, *,**}\\
 Published Online **** 2016 in SciRes.
 \href{http://www.scirp.org/journal/*****}{\color{blue}{\underline{\smash{http://www.scirp.org/journal/****}}}} \\
 \href{http://dx.doi.org/10.4236/****.2016.*****}{\color{blue}{\underline{\smash{http://dx.doi.org/10.4236/****.2016.*****}}}} \\
 \et}
 \fancyhead[R]{\includegraphics{pic1.ps}}

 $\mbox{ }$

 \vskip 12mm

{ 

{\noindent{\huge\bf\color{myaqua}
  Entropy of Causal Horizons}}
%
\\[6mm]
{\large\bf Eric M Howard$^1$}}
\\[2mm]
{ 
 $^1$Department of Physics and Astronomy, Macquarie University\\
Email: \href{mailto:eric.howard@mq.edu.au}{\color{blue}{\underline{\smash{eric.howard@mq.edu.au}}}}
 \\[4mm]
Received **** 2016
 \\[4mm]
Copyright \copyright \ 2016 by author(s) and Scientific Research Publishing Inc. \\
This work is licensed under the Creative Commons Attribution International License (CC BY). \\
\href{http://creativecommons.org/licenses/by/4.0/}{\color{blue}{\underline{\smash{http://creativecommons.org/licenses/by/4.0/}}}}\\
 \includegraphics{pic2.ps}

\lin{5}{7}

 { 
 {\noindent{\large\bf\color{myaqua} Abstract}{\bf \\[3mm]
 \textup{We analyze spacetimes with horizons and study the thermodynamic aspects of causal horizons, suggesting
that the resemblance between gravitational and thermodynamic systems has a deeper quantum mechanical origin. We find that the observer dependence of such horizons is a direct consequence of associating a temperature and entropy to a spacetime. The geometrical picture of a horizon acting as a one-way membrane for information flow can be accepted as a natural interpretation of assigning a quantum field theory to a spacetime with boundary, ultimately leading to a close connection with thermodynamics.
}}}
   \\[4mm]
 {\noindent{\large\bf\color{myaqua} Keywords}{\bf \\[3mm]
 black hole entropy; causal horizon;black hole horizon;
 black hole thermodynamics; Rindler spacetime;
 Rindler horizon; quantum gravity; thermodynamics of spacetime
}

 \fancyfoot[L]{{\noindent{\color{myaqua}{\bf How to cite this
 paper:}} Eric M Howard (2016)
 Entropy of Causal Horizons.
 ***********,*,***-***}}

\lin{3}{1}

\sectionn{Introduction}

The interaction between gravity and thermodynamics has become a subject
of growing interest since the discovery of Bekenstein-Hawking entropy.\cite{Bek73}
The analogy between black hole mechanics and thermodynamics led Bekenstein \cite{Erice} and Hawking to argue that
black holes should be viewed as true thermodynamic systems, characterised by entropy and temperature.\cite{Bk1}

Several authors have worked on extending black hole entropy to the more general case where
the connection between entropy and area would generalize to any horizon, like accelerated horizons in Rindler spacetime.
One intriguing conclusion was the idea that horizon entropy arises in a natural way just because gravity is
emergent from thermodynamics. The new perspective motivated Jacobson,\cite{TJ1}\cite{TJ3} Padmanabhan \cite{tpcqg} \cite{tpapoorva} and Verlinde to suggest an interpretation of gravity as a thermodynamic phenomenon.\cite{cqgpap}

While the concepts of temperature and entropy remain associated to spacetimes with horizons,
a theory that fully explains the quantum structure of spacetime should ultimately remain
independent of any microscopic description of the horizons themselves. The hard part here is the lack
of understanding about the nature of the microstates whose counting would produce the horizon entropy, while
the geometrical picture \cite{howard2} of the black hole entropy would still remain a good approximation for the theory.

The notion of entropy is closely connected to the accessibility of information, therefore an observer dependent concept.
If a geometrical interpretation of gravity is accepted, surfaces acting as one-way membranes 
for information will exist, leading to a deeper connection with thermodynamics. 
An inevitable connection between one-way membranes that arise in a spacetime with horizons and thermodynamical entropy is already obvious here.\cite{thorne}

\renewcommand{\headrulewidth}{0.5pt}
\renewcommand{\footrulewidth}{0pt}

 \pagestyle{fancy}
 \fancyfoot{}
 \fancyhead{} 
 \fancyhf{}
 \fancyhead[RO]{\leavevmode \put(-90,0){\color{myaqua}Eric M Howard} \boxx{15}{-10}{10}{50}{15} }
 \fancyhead[LE]{\leavevmode \put(0,0){\color{myaqua}Eric M Howard}  \boxx{-45}{-10}{10}{50}{15} }
 \fancyfoot[C]{\leavevmode
 \put(0,0){\color{lightaqua}\circle*{34}}
 \put(0,0){\color{myaqua}\circle{34}}
 \put(-2.5,-3){\color{myaqua}\thepage}}

 \renewcommand{\headrule}{\hbox to\headwidth{\color{myaqua}\leaders\hrule height \headrulewidth\hfill}}

The interesting fact is that so far we have knowledge only about two types of physical processes that can
lead to distinct regions of spacetime, some of them completely disconnected from each other so that information from
one spacetime patch will not be accessible to an observer located within another patch: light and gravity. 
Therefore, some strong connection between light, gravity and thermodynamics must exist. The AdS/CFT conjecture is
intimately connected to this argument. 

The extreme conditions for quantum, relativistic and gravitational regimes: the very small, the very heavy and the very fast, seem to intersect at a fundamental level, leading to a more general principle
that so far seems to have eluded our complete understanding. A theory formulating a full description of the quantum structure of the spacetime should be responsible to explain all three extreme conditions and in this context, should involve themodynamics principles in its basic arguments.

If the connection between thermodynamics and spacetimes with horizons has a
fundamental meaning, the argument should be able to be extended to
any general horizon, therefore we can assign entropy to any horizon, 
irrespective of whether the horizons are observer independent (black holes) or observer and coordinate dependent (de Sitter and Rindler horizons). 

Our argument here is that the entropy of spacetimes arise because
there is information hidden behind the horizon, generating different spacetime regions, causally disconnected by an entangling null surface acting as a one-way information barrier. An exploration of the Quantum Field Theory of such a spacetime should suggest a natural association with a temperature. The inertial observer will ascribe the entropy given by the logarithm of the number of internal states. Since a horizon is capable to block information from an observer, 
it is feasible to associate an entropy with the horizon. Here we show that the entropy of an 
event in flat empty space depends on the observer that measures it. We argue that this dependence originates in
a pure quantum mechanical effect of the fact that for an accelerated observer, the event is
hidden behind the Rindler horizon.

\sectionn{Horizon entropy and observers}

\label{gen}

The idea of entanglement entropy has
emerged as a measure of entanglement in order to explain the internal correlations within a quantum system.
In quantum  field theory the entanglement between a spatial region on
a fixed Cauchy surface region and its complement,
is found to be divergent because of the presence of short range correlations near the
entangling surface that separates the spacetime regions.\cite{birrel}  

Entropy can be viewed as a measure of missing or unavailable
information about a system. It would be expected that some entropy to be associated with the event
horizon, since it hides information.
The horizon has an ``entanglement entropy'' associated with the quantum field nearby, produced by
tracing out its interior inaccessible modes and ultimately correlated with the
exterior accessible (to the observer) modes of the field.  
For a fixed background spacetime continuum, the entanglement entropy is infinite and a short distance cut-off
on the field degrees of freedom needs to be introduced in order to obtain a finite entropy 
If the cut-off is near the Planck scale, the entropy is
of the same order of magnitude as the horizon itself.

The divergence of the entanglement entropy requires an ultraviolet cut-off at Planck scale.
In order to regulate this divergence, gravity should be treated as a quantum field.
As gravity is non-renormalizable, the entanglement entropy should also be non-renormalizable.
Furthermore, a complete theory of quantum gravity should explain the cut-off and solve the divergence of the entanglement
entropy together with the ultraviolet problem. \cite{Bombelli} 

Sorkin \cite{chb} \cite{tpe} has proposed the idea that entanglement entropy offers a microscopic explanation for
Bekenstein-Hawking entropy. In Quantum field theory, space can be partitioned
into subregions. The degrees of freedom can be isolated inside the horizon from the surrounding spacetime, 
allowing a state on the Cauchy surface to be a pure state like the vacuum, whereas the region of spacetime outside the horizon may contain mixed states such as a thermal states of non-zero temperature.

In this way, the horizon will posses entropy, while offering a clear division of degrees of freedom outside the
horizon that are observable, and degrees of freedom behind the horizon that are not perceived by any observer.

At a conceptual level, the entanglement entropy does not represent the ignorance of the observer about state of
the spacetime region behind the horizon but a measure of the observer's ignorance
about the state outside the horizon that is due to the observer's inability
to measure the states behind the horizon. The increase of entropy on the horizon
is a effect of the entanglement as well as simple relativistic causality. 

The mathematical form of the association of temperature with the horizon is universal and it does not distinguish between different types of horizons (Rindler horizon in flat space, Schwarzschild black hole event horizon, de Sitter horizon). The fact that the temperature and entropy both arise in the same way, suggest a strong direct association of entropy with all horizons. \cite{Bousso}

We extend this concept to Rindler horizons or any type of horizons that exhibit causal order.
We know that Rindler coordinates describe an observer moving along a constant acceleration path.\cite{UW} In Minkowski
spacetime, Rindler coordinates depict the near-horizon limit of a black hole. In AdS spacetime, Rindler coordinates 
divide AdS in two causally disconnected regions, known as left and right (in causal contact with the observer) Rindler wedges. Thermodynamical entropy is here produced by variables that are observable (energy of
a system), whereas other variables (exact positions of particles) remain unobservable.

As a consequence, the entropy ascribed to a localized system by different observers is not necessarily the same. 
The entropy of an event in flat empty space will be differently ascribed by an
inertial and a Rindler observer. 
The entropy seen by a Rindler observer should have a
bound given by by $E_{Rindler} /T$, where $E_{Rindler}$ is the 
energy of an object associated with an event observed by the Rindler observer and $T$ the
temperature measured by the observer.

For an inertial observer, the vacuum state is given by
the Minkowski vacuum  $|0\rangle$. The
excitation of the vacuum that could be due to the presence of an object
is given by the density matrix
\begin{equation}
\rho = {1 \over n} \sum_{i=1}^n |1;i \rangle \langle1;i|,
\end{equation}
The perturbation of the vacuum state is 
in an undetermined microstate. The $|1;i\rangle$ are Minkowski states
represented by the $n$ possible microstates of the perturbation.

The measured entropy will also contain
an additional contribution introduced by the acceleration of the Rindler observer. \cite{UW2} 
The measured energy $E_{Rindler} = \delta E$ is given by the difference in
Killing energies when the perturbation is present and when it is not. 
The entropy of each state will be divergent.  
The Rindler object entropy $S_{Rindler}$ is the similar difference $\delta S$ between the two
entropies in the right Rindler wedge, in causal contact with the observer.

The entropy as measured by an inertial observer is given by
\begin{equation}
S = - \mbox{Tr} \rho \ln \rho = \ln n.
\end{equation}
with $\rho$ being the 
density matrix $\rho_{Rindler}$ equivalent for a Rindler
observer.

For a perturbation with the energy $\delta E$, the density matrix $\rho_{R1}$ for a Rindler observer can be calculated by integrating the unobservable modes.
The entropy will become $S_1 = - {\rm Tr}\ (\rho_{R1} \ln \rho_{R1})$.
The inertial vacuum state density matrix in Rindler frame is $\rho_{R0}$ with entropy $S_0 = - {\rm Tr}\ (\rho_{01} \ln \rho_{R0})$. 

For the Rindler observer, the difference in entropies $\delta S = S_1 - S_0$ is the entropy of the vacuum state. Near the horizon, for $\kappa \to \infty$ we have
  \begin{equation}
\delta S = \beta \delta E = \frac{2\pi}{\kappa} \delta E
\label{delS}
\end{equation}
where $\beta=T^{-1}$.

The Rindler observer will not measure the same entropy of the perturbation.
Following Sorkin hypothesis for black hole entropy and apply to all causal horizons, \cite{howard1} the increase of entanglement entropy is a direct consequence of the fact that a Rindler horizon is a causal barrier that separates two causally disconnected regions of spacetime. The whole spacetime is visible to the inertial observer, therefore in causal contact with the observer. 

The entropy measured by a Rindler observer is given by the right Rindler wedge 
in causal contact with the observer, when the object crosses the Rindler horizon. 
The entropies associated with this perturbation
by a Minkowski and Rindler
observer behave in different ways.

If we trace the Minkowski vacuum $|0\rangle \langle 0|$ 
over the left Rindler wedge, the thermal density matrix $\rho_{R0}$ will
contain all the information that a Rindler observer can
access in the Minkowski vacuum:
\begin{equation}
\rho_{R0} = \mbox{Tr}_{left} |0 \rangle \langle 0 |.
\end{equation}
We consider a new density matrix $\rho_{R1}$ corresponding to the Rindler 
description of the state $\rho_{Rindler}$,  containing information about the object added to
the Minkowski vacuum $|0 \rangle$ and further compare it to $\rho_{R0}$. 
The density matrix $\rho_{R0}$ represents the thermal
equilibrium.
The new density matrix $\rho_{R1}$ is
\begin{equation}
\rho_{R1} = \mbox{Tr}_{left} \rho_{Rindler} = \mbox{Tr}_{left}
\frac{1}{n} \sum_{i=1}^n |1;i \rangle \langle1;i|.
\end{equation}

If $\rho_{R1}=\rho_{R0} + \delta \rho$, for
$\kappa \to \infty$ we only take into account states with $\delta \rho/\rho_{R0}\ll 1$.
The difference in entropy is
\begin{eqnarray}
-\delta S &=& {\rm Tr}\ (\rho_{R1} \ln \rho_{R1}) - {\rm Tr}\ (\rho_{R0} \ln \rho_{R0}) 
\simeq {\rm Tr}\ (\delta \rho \ln \rho_{R0})\nonumber\\
&=& {\rm Tr}\ (\delta \rho (-\beta H)) = - \beta {\rm Tr}\ \left((\rho_{R1} -\rho_{R0})H\right) \equiv -\beta \delta E
\end{eqnarray} 
with Tr $\delta \rho \approx 0$ and
$\rho_0 =Z^{-1}\exp(-\beta H)$. Here $H$ is the Hamiltonian in Rindler frame. 
The difference $\delta E$ is defined as a difference in 
expectation values of the Rindler Hamiltonian between the two states.

If the perturbation on the Rindler density matrix
$\rho_{R0}$ is very small, we have the new density matrix
$\rho_{R1} = \rho_{R0} + \delta \rho$, with ``$\delta
\rho \ll \rho_{R0}$''. In the limit of large numbers $n e^{-E/T} \gg 1$, the approximation still holds.  

The difference in entropies $\delta S$ using a Taylor expansion $\rho_{R0}$ is
\begin{equation}
\label{Sresult}
\delta S \approx - \mbox{Tr}\left[ \delta \rho {\delta (\rho \ln \rho)
    \over \delta \rho}\Big|_{\rho=\rho_{R0}} \right] = - \mbox{Tr} [
    \delta \rho (1 + \ln \rho_{R0})] = - \mbox{Tr} [\delta \rho \ln
    \rho_{R0}],
\end{equation}
with $\mbox{Tr} \rho_{R1} =
\mbox{Tr} \rho_{R0}$. 

As the initial density matrix is
thermal, we have $\rho_{R0} = e^{-H/T}/(\mbox{Tr} e^{-H/T})$ and
\begin{equation}
\delta S \approx - \mbox{Tr} [ \delta \rho (-H/T)] = {\mbox{Tr} [H
    (\rho_{R1} - \rho_{R0})] \over T} = {\delta E \over T},
\end{equation}
with $\mbox{Tr}(\delta \rho)=0$. 

If by introducing a perturbation, the energy increases by $\delta E$, but the entropy can't be increased by more than the amount of thermal energy, independent of the nature of the excitation. This energy bound arises from tracing over the left Rindler
wedge, being a direct consequence of the existence of the
horizon itself, as a causal barrier between spacetime regions.

The energy bound will be given by
\begin{equation} \label{enbound}
\delta S < \delta S_{max} = {\delta E \over T}.
\end{equation}

For a Rindler observer, when quantum mechanical effects are taken into account, the entropy and energy are linked by the Bekenstein bound, associated with thermodynamics, the holographic principle and the covariant entropy bound of quantum gravity. Bekenstein conjectured the entropy bound to be a universally valid principle for complete systems
in nature. Using quantum statistics, Bekenstein showed that the entropy bound
holds for complete systems with non-interacting massless quantum fields in flat spacetime, idea criticized by several authors (Unwin, Page, Unruh, Wald, Pelath). 

If a system violates this bound (highly entropic object), for a perturbation
with large number of internal states, described by a large entropy with fixed energy,
the entropy and energy appear independent to an inertial observer. The interesting fact is that this bound does not depend on the gravitational constant, suggesting a geometric nature of the horizon entropy and ultimately an emergent origin of gravity.

An observer that falls into a black hole will have access to a different amount of information compared to another observer that is stationary outside the horizon. Similarly, a Rindler observer will have access to different regions of spacetime in comparison to the inertial observer. The inertial observer measures zero temperature and entropy for the vacuum state, whereas a Rindler observer measures a finite temperature and non-zero divergent entropy. Entropy and temperature are apparently observer dependent, due to the effects of the quantum field on the horizon. 
 
If the field configuration of the vacuum state behind the horizon
is traced over, a density matrix $\rho$ for the field configuration outside the horizon can be detected. The behaviour of the horizon as a one way membrane blocking all the information to the outside observer suggests an observer dependent entropy that can be associated with any causal horizon.\cite{reftwo} The entropy arises from quantum cross-correlations which exist across the horizon. The quantum state of the horizon is a correlated state with a component inside the horizon and another one outside. If the observer traces over the states behind the horizon, the spacetime outside the horizon will be given by a density matrix.\cite{Solo} The spacetime will contain two isolated regions, separated by the surface horizon whereas the description of each spacetime region will be given by tracing over the other complementary region. The thermal density matrix and therefore the temperature of the horizon can be obtained from the integration of modes that are hidden by the horizon. 

The increase in entropy is due to a separate contribution originated in the quantum field within the spacetime near the horizon, concept closely related to the “entanglement entropy” generated by correlations between quantum fields on each side of the horizon. The entropy  measures the lack of information about the measurement outcomes implied by the Heisenberg uncertainty relation and at the same time measures the amount of entanglement between the two sides of the horizon. In other words, the von Neumann entropy describes the observer statistical ignorance of the specific microstate (a translation of the classical entropy of Gibbs and Boltzmann) while expressing the amount of entanglement.\cite{gravitation} This idea supports a possible equivalence between interpretations of von Neumann entanglement entropy and Boltzmann statistical entropy concepts.

An analysis of the thermodynamics across a horizon discloses a strong relation between thermality and von Neumann entropy describing entanglement. The observer interacts with the quantum field observables defining the quantum state as a density matrix. The entropy will quantify the uncertainty, due to entanglement of measurements of the quantum field seen by an accelerated observer. S. Lloyd has already proposed the idea that quantum uncertainty gives rise to entanglement, showing that the Heisenberg uncertainty principle and quantum entanglement and are two inextricably connected phenomena. Jacobson derived Einstein's equations from the vacuum entanglement entropy, on the understanding of a connection between the semi-classical Einstein equations and a maximal vacuum entanglement hypothesis.\cite{dawoodnew} 

If the horizon entropy is given by the Bekenstein-Hawking formula, then the second law of thermodynamics leads to Einstein's equations. More recently, Ryu-Takayanagi have shown that there is a deeper relationship between entanglement and gravity in the AdS/CFT conjecture, where the entanglement entropy between a CFT patch and its neighbourhood is given a minimal AdS area with the same boundary as the patch.  In the context of the AdS/CFT correspondence, entanglement entropy is defined on the AdS boundary and proportional to the area of a minimal surface in the bulk of the AdS spacetime.

In Rindler spacetime, the observer will have access to the right Rindler wedge and observe in a natural way all the thermal fluctuations, as the observer attributes a density matrix to a quantum state only after integrating out the unobservable modes. Here, the thermal fluctuations have a quantum mechanical origin, suggesting an artificial distinction between quantum and thermal properties applied to the horizon that should disappear in a correct complete theory of quantum gravity. 

As a direct consequence, any matter plunging towards the horizon will take an infinite amount of time in order to cross the horizon for the outside observer.\cite{WaldRev} On the other hand, the quantum effects will smear the location of the horizon. Any description of the location of the horizon has to take into account these quantum fluctuations. The physics loses its predictive power about what happens to the information encoded at the horizon. \cite{Frolov} Both location of the horizon and the crossing time are unpredictable to the outside observer.

The Rindler observer (or an observer outside a black hole horizon) attributes a finite temperature to the inertial vacuum
and associates an entropy change to a horizon while the inertial observer (or an observer falling through the horizon) sees none of these phenomena. 

The entropy of matter crossing the horizon is $\delta S = \delta E/T_{\rm horizon}$ where the horizon plays the role of a causal barrier for the Rindler observer, with inaccessible internal degrees of freedom and temperature $T_{\rm horizon}$. 

The concept of entropy is associated with the missing information
but this does not lead to non zero entropy, leading to the ``information loss problem''. \cite{carlip}
The horizon hides information only from the point of view of an external observer. 
From this perspective, the Rindler entropy is observer dependent.

The connection between entanglement entropies and areas of extremal surfaces leads us to the conclusion that the thermodynamic properties of a system need to be conceived as observer dependent, depending on the spacetime region where the system is perceived. The entanglement entropy  for  a  quantum field  theory will obey the  area law. The quantum correlations between the interior and the exterior of the horizon will account for the  black hole entropy.

Entanglement entropy was already proposed as a probe of the architecture of the spacetime in quantum gravity.
The point here is that the quantum source of black hole entropy is associated with the thermal behaviour of the spacetime,
suggesting a deeper connection between entanglement and thermodynamics. \cite{ted} As a consequence, a consistent and complete theory of quantum gravity should be built in a formalism using “histories” and “path integral” formulations of quantum mechanics rather than the spatial (“3 + 1”) canonical quantization. In fact, the area law itself, that gives a geometric character to entropy, brings the unexpected link between gravity, spacetime geometry, thermodynamics and quantum  field theory. The horizon will play the role of an entangling membrane that separates the degrees of freedom between the the two sides of a black hole. \cite{BFZ}

This idea is also backing up induced gravity models like the eternal AdS black hole, in AdS/CFT conjecture, with the horizon entropy being defined as the entanglement entropy between the microscopic degrees of freedom of the boundary CFT and its thermofield double. \cite{Ray} The area law is also confirmed by loop quantum gravity \cite{Sahlmann} and spin foam theory, \cite{cov} with horizon entropy defined by the entanglement entropy of spin-network connections crossing an entangling surface.

\sectionn{Conclusions}\label{sec:conclusions}

The paradigm described in this paper, relying on a study of general causal horizons,\cite{howard3} is that spacetime itself (not just the black hole interior or the event horizon) has entropy and can be described in geometric, macroscopic terms (area law) and thermodynamic concepts, as a manifestation of a deeper quantum nature of spacetime that at this time eludes us. The obvious consequence of associating the entropy of causal horizons to the missing or hidden information is the fact that the entropy is proportional to horizon area, with respect to the observer that perceives the horizon.

The thermodynamic route to gravity brings many conceptual ramifications. In such a paradigm where the entanglement of quantum fields can be considered the source of entropy and has a geometrical nature,\cite{Replica} any consistent quantum gravity theory should take into account the thermodynamic properties of spacetime. Entropy encodes all information about the dynamics of spacetime and finding its extreme will lead to Einstein field equations, as Jacobson demonstrated. In a thermodynamic description, the presence of entropy leads to equipartition of energy among the degrees of freedom and consistent field equations. \cite{Bek03}

An observer plunging into a black hole will not perceive its event horizon in the same way as an observer orbiting the black hole. Therefore observers in different states of motion will have access to distinct patches of spacetime. The plunging observer measures distinct features compared to the observer near the black hole. At first glance, entropy appears to be an observer dependent notion. At the same time, each observer is located in a different region of the spacetime and will measure the properties of the horizon, from within a distinct spacetime patch and relative to the quantum field associated with that region. A local Rindler observer perceives a local causal horizon together with its thermal behaviour associated to an event. A connection between quantum fluctuations and thermal fluctuations must exist here. 

This hypothesis is intimately related to the idea that the field equations,\cite{landau100} therefore gravitational dynamics are associated to horizon thermodynamics. For a spacetime with horizons, the quantum field theory for this spacetime will possess a horizon with the properties of a black body of finite temperature. A connection between horizon thermality and gravity becomes obvious here.

From a different perspective, the emergence of classical spacetime continuum \cite{tHoo} seems directly related to the quantum entanglement of degrees of freedom in a non-perturbative description of quantum gravity.  \cite{LSuss} \cite{FiSsuss}  A good understanding of how classical spacetime emerges from an underlying quantum system \cite{hawking000}would give us insight into the full quantum nature of spacetime.

 {\color{myaqua}

\end{document}